\documentclass[aps,prc,preprint,showpacs,superscriptaddress]{revtex4}
\usepackage{epsfig}
\usepackage{amsmath}
\usepackage[hyperref]{hyperref}

\begin{document}

\title{Deeply Virtual Compton Scattering on nucleons and nuclei
in generalized vector meson dominance model}

\author{K. Goeke}
\email{Klaus.Goeke@tp2.rub.de}
\affiliation{Institut f\"ur Theoretische Physik II, Ruhr-Universit\"at-Bochum, D-44780 Bochum, Germany}

\author{V. Guzey}
\email{vguzey@jlab.org}
\affiliation{Theory Center, Jefferson Lab, Newport News, VA 23606, USA}

\author{M. Siddikov}
\email{Marat.Siddikov@tp2.rub.de}
\affiliation{Institut f\"ur Theoretische Physik II, Ruhr-Universit\"at-Bochum, D-44780 Bochum, Germany}
\affiliation{Theoretical  Physics Dept,Uzbekistan National University, Tashkent 700174, Uzbekistan}

\preprint{JLAB-THY-08-774}

\begin{abstract}

We consider Deeply Virtual Compton Scattering (DVCS) on
nucleons and nuclei in the framework of generalized vector
meson dominance (GVMD) model.
We demonstrate that the GVMD model provides a good description 
of the HERA data on the dependence of the proton DVCS cross section 
on $Q^2$, $W$ (at $Q^2=4$ GeV$^2$) and $t$. At $Q^2 = 8$ GeV$^2$, 
the soft $W$-behavior of the GVMD model somewhat underestimates 
the $W$-dependence of the DVCS cross section due to the hard contribution
not present in the GVMD model.
We estimate  $1/Q^2$ power-suppressed corrections 
to the DVCS amplitude and the DVCS cross section
and find them large.
We also make predictions for the nuclear DVCS amplitude and cross section in the 
kinematics of the future Electron-Ion Collider.
We predict significant nuclear shadowing, which matches well predictions 
of the leading-twist nuclear shadowing in DIS on nuclei.

\end{abstract}

\maketitle

\section{Introduction}
\label{sec:Intro} 

During the last decade, one of main focuses of hadronic physics has been the
study of the hadronic structure using hard exclusive reactions,
such as  Deeply Virtual Compton Scattering (DVCS),
$\gamma^{\ast}N \to \gamma N$, and hard exclusive meson production (HEMP),
$\gamma^{\ast}N \to M N$.
These processes have been the subject of intensive theoretical and
experimental investigations~\cite{Mueller:1998fv,Ji:1996nm,Ji:1998pc,Radyushkin:1996nd,Radyushkin:1997ki,Radyushkin:2000uy,Collins:1998be,Collins:1996fb,Brodsky:1994kf,Goeke:2001tz,Diehl:2000xz,Belitsky:2001ns,Diehl:2003ny,Belitsky:2005qn}.
In addition, there were investigated ''inverse'' hard exclusive reactions 
such as 
$\gamma N \to \gamma^{\ast}N \to l^{+}l^{-}N$~\cite{Berger:2001xd}
 and
$\pi N \to \gamma^{\ast}N \to l^{+}l^{-}N$~\cite{Berger:2001zn}, and
''$u$-channel'' reactions such as $\gamma^{\ast} \gamma \to \pi \pi$~\cite{Diehl:2000uv}.

The interest to the DVCS and HEMP reactions is motivated by the fact that 
in the Bjorken limit (large $Q^2$), the corresponding amplitudes factorize~\cite{Collins:1998be,Collins:1996fb}
in convolution of perturbative (hard) coefficient
functions with nonperturbative (soft) matrix elements, 
which are parameterized in terms of generalized parton distributions 
(GPDs).
GPDs are universal (process-independent) functions that contain information
on parton distributions, form factors and correlations in hadrons. 
GPDs also parameterize parton correlations in
matrix elements
describing transitions between two different hadrons, which appear in 
reactions such as
e.g.~$\gamma^{\ast}p \to \pi^{+}n$~\cite{Mankiewicz:1998kg,Frankfurt:1999fp}.

While the description of DVCS and HEMP based on the factorization approach is
most general, 
in experiments the values of the virtualities $Q^2$ 
are below the range
required for the validity
of the factorization theorem~\cite{Airapetian:2001yk,Airapetian:2006zr,Ye:2006gza,Stepanyan:2001sm,Munoz Camacho:2006hx,Girod:2007jq}.
Hence, contributions of higher-twist
effects might be substantial (it is an open issue how large these
effects are), which will affect the extraction of 
GPDs from the data. 
Therefore, it is important to have an effective model for the DVCS and HEMP 
amplitudes, which would interpolate between the photoproduction ($Q^2 \approx 0$)
and deep inelastic ($Q^2 \sim {\cal O}({10})$ GeV$^2$) regimes.

In this paper, using the generalized vector meson dominance (GVMD) 
model~\cite{Fraas:1974gh,Ditsas:1975vd,Shaw:1993gx},
which is consistent with perturbative QCD at small 
transverse distances~\cite{Frankfurt:1997zk},
we derive expressions for the amplitudes of DVCS on nucleons and nuclei, which  
are valid at high energies and which are applicable over a wide range of $Q^2$.
We show that the resulting cross section of DVCS on nucleons compares well
to the HERA data~\cite{Chekanov:2003ya,Aktas:2005ty}.
In particular, the dependence of the DVCS cross section on $Q^2$, $W$ (at  $Q^2$=4 GeV$^2$) and $t$ are reproduced rather well; the $W$-dependence of the 
cross section at  $Q^2$=8 GeV$^2$ is somewhat underestimated, which 
can be interpreted as due to the onset of the 
hard regime beyond the soft dynamics of the GVMD model.

We also estimate the relative contribution of $1/Q^2$-corrections, 
which correspond to the higher-twist corrections in perturbative QCD~\cite{Kivel:2000fg,Freund:2003qs,Radyushkin:2000ap}. 
We show that these corrections are large:
the contribution of the $1/Q^2$-corrections 
to the DVCS amplitude at 
$t=t_{{\min}}$ is 20\% at $Q^2=2$ GeV$^2$, 11\% at $Q^2=4$ GeV$^2$
and 6\% at $Q^2=8$ GeV$^2$;
the contribution of the $1/Q^2$-corrections 
 to the $t$-integrated DVCS cross
section is 56\% at $Q^2=2$ GeV$^2$, 32\% at $Q^2=4$ GeV$^2$
and 17\% at $Q^2=8$ GeV$^2$.

We also make predictions for the DVCS cross section on nuclear targets, which are relevant for the physics program of the future Electron-Ion Collider.
We predict significant nuclear shadowing, which matches well predictions 
of the leading-twist nuclear shadowing in DIS on 
nuclei~\cite{Frankfurt:2003zd}.

The hypothesis of vector meson dominance (VMD)~\cite{Feynman} assumes 
a definite relation between
the amplitude of the photon (real or virtual)-hadron interaction,
${\cal A}(\gamma_{{\rm tr}}^{\ast}+T \to \dots)$,
 and a linear 
combination of the amplitudes of the corresponding strong production by
transversely polarized vector mesons,  ${\cal A}(V_{{\rm tr}}+T \to \dots)$,
\begin{equation}
{\cal A}(\gamma_{{\rm tr}}^{\ast}+T \to \dots)=\sum_{V=\rho, \omega,\phi} \frac{e}{f_V} \frac{m_V^2}{m_V^2+Q^2} {\cal A}(V_{{\rm tr}}+T \to \dots) \,,
\label{eq:vmd_1}
\end{equation}
where $f_V$ is the coupling constant determined from the
$V \to e^{+} e^{-}$ decay; $m_V$ is the vector meson mass;
$Q^2$ is the virtuality of the photon; $T$ denotes any
hadronic target. Note that Eq.~(\ref{eq:vmd_1}) is written for the transversely
polarized photons.
 In Eq.~(\ref{eq:vmd_1}), we took into account only
the contribution of the $\rho^0$, $\omega$ and $\phi$ mesons.

The VMD model and its generalizations explain a large 
wealth of data on the real and virtual ($Q^2 < 1$ GeV$^2$) photon-hadron scattering,
which include the pion electric form factor, total cross sections of 
photon-nucleon and  photon-nucleus scattering (inclusive structure functions),
exclusive production of
vector mesons on nucleons and nuclei, 
 exclusive production of pseudoscalar mesons, 
for a review, see~\cite{Bauer:1977iq}. 

As the virtuality of the photon increases, $Q^2 > 1$ GeV$^2$, the simple 
VMD model, see Eq.~(\ref{eq:vmd_1}), becomes inadequate since
it leads to the violation of the approximate Bjorken scaling.
In order to restore the approximate Bjorken scaling, the simple VMD model
can be generalized~\cite{Fujikawa:1972ux}. This can be done using 
the model-independent method of mass-dispersion representation 
for the virtual photon-hadron scattering amplitude~\cite{Gribov:1968gs}.

In order to illustrate the approach, let us 
consider the forward virtual photon-hadron scattering amplitude.
The dispersion representation for the imaginary part of 
${\cal A}(\gamma^{\ast}_{{\rm tr}}+T \to \gamma^{\ast}_{{\rm tr}}+T)$ reads
\begin{equation} 
\Im m \,{\cal A}(\gamma_{{\rm tr}}^{\ast}+T \to \gamma_{{\rm tr}}^{\ast}+T)_{|t=0}=\int \frac{dM^2 M^2}{M^2+Q^2} \frac{dM^{\prime 2} M^{\prime 2}}{M^{\prime 2}+Q^2} \frac{e}{f_V} \sigma_{V V^{\prime}} \frac{e}{f_{V^{\prime}}} \,,
\label{eq:vmd_1b}
\end{equation}
where $\sigma_{V V^{\prime}}$ is the $V +T \to V^{\prime}+T$ scattering cross section
(spectral function)
which weakly depends on the masses $M$ and $M^{\prime}$.
The main idea of the GVMD 
model~\cite{Fraas:1974gh,Ditsas:1975vd,Shaw:1993gx} is to approximate 
Eq.~(\ref{eq:vmd_1b}) by an infinite series of (ficticious) vector mesons of
ever increasing mass, allowing for both  diagonal ($V +T \to V+T$) and non-diagonal
($V +T \to V^{\prime}+T$) transitions. The role of the non-diagonal 
transitions is to partially cancel the diagonal transitions
so that, effectively,  $\sigma_{V V^{\prime}} \propto 1/M^2$ for large $M^2$.
This softens the spectral function and 
leads to  the approximate Bjorken scaling of the inclusive
structure function $F_2(x,Q^2)$, see \cite{Frankfurt:1997zk} for the discussion.

In the language of the color dipole model, the fact that $\sigma_{V V^{\prime}} \propto 1/M^2$ for large $M^2$ means that besides dipoles of large
transverse sizes, the virtual photon also contains small transverse-size
dipoles. The latter fact is called color transparency.

One should note that, 
while the simple vector meson dominance model fails to reproduce the approximate 
scaling of the {\it inclusive} structure function $F_2(x,Q^2)$ (see above),
the simple VMD model predicts the correct
$Q^2$-behavior of cross sections of {\it exclusive} reactions, such as
$\gamma^{\ast} p \to \pi^{+}n$~\cite{Fraas:1971hk,Dar:1971wh}.
This is also true for DVCS: Even the simple VMD model provides the correct
$Q^2$-behavior of the DVCS cross section (up to logarithmic corrections).

The structure of this paper is the following. 
In Sect.~\ref{sec:dvcs-nucleon}, we explain main assumptions of the GVMD model,
which we further generalize to take into account a non-zero momentum transfer $t \neq 0$.
We derive the expression for the amplitude of DVCS on the nucleon and make predictions
for the DVCS cross section.
We demonstrate that the GVMD model provides a good description of the HERA data on the
$W$, $Q^2$ and $t$-dependence of the cross section 
of DVCS on the proton~\cite{Chekanov:2003ya,Aktas:2005ty}.
In this section, we also estimate $1/Q^2$-corrections to the DVCS amplitude and the 
DVCS cross section.
Predictions for the nuclear DVCS amplitude and for the cross section of DVCS on nuclei
in the collider kinematics are presented in Sect.~\ref{sec:dvcs-nucleus}.
In Sect.~\ref{sec:conclusions}, we summarize and discuss our results.

\section{DVCS on the nucleon}
\label{sec:dvcs-nucleon}

In this section, we extend the generalized vector meson dominance (GVMD) model~\cite{Fraas:1974gh,Ditsas:1975vd,Shaw:1993gx}
to the off-forward case and apply it
 to Deeply Virtual Compton 
Scattering (DVCS) on the nucleon.

\subsection{DVCS amplitude}

\begin{figure}[t]
\begin{center}
\epsfig{file=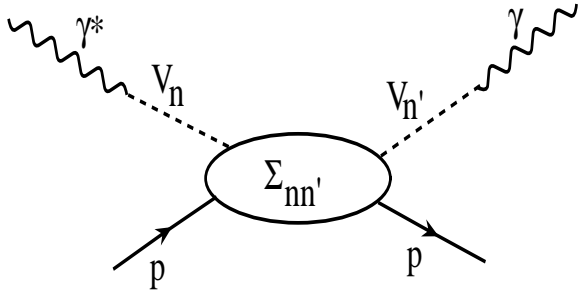,width=12cm,height=8cm}
\vskip -1cm
\caption{The DVCS amplitude in the GVMD model, see Eq.~(\ref{eq:gvmd_1}).}
\label{fig:dvcs_gvmd}
\end{center}
\end{figure}

 The GVMD model assumes that the virtual 
(real) photon interacts with the hadronic target
by fluctuating into a coherent and infinite sum 
of ficticious vector mesons $V_n$. 
Then, the DVCS amplitude at the photon level,
${\cal A}(\gamma^{\ast} p \to \gamma p)$, can be graphically presented as depicted in Fig.~\ref{fig:dvcs_gvmd}.

In the GVMD model, the DVCS amplitude for transversely polarized virtual
photons reads
\begin{equation}
{\cal A}(\gamma_{{\rm tr}}^{\ast}+p \to \gamma+ p)=\sum_{n,n^{\prime}=0}^{\infty} \frac{e}{f_n}
\frac{M_n^2}{M_n^2+Q^2} \Sigma_{n, n^{\prime}}(W,t)
\frac{e}{f_{n^{\prime}}}  \,,
\label{eq:gvmd_1}
\end{equation} 
where $W^2=(p_{\gamma^{\ast}}+p)^2$ with $p_{\gamma^{\ast}}$ the momentum of the
initial photon and $p$ the momentum of the initial proton;
$t=(p^{\prime}-p)^2$ with $p^{\prime}$ the momentum of the final proton.
The masses $M_n$ and the coupling constants $f_n$ are connected by the following
relations,
\begin{equation}
\frac{M_n^2}{M_0^2}=\frac{f_n^2}{f_0^2}=(1+2n) \,, 
\label{eq:gvmd_2}
\end{equation} 
where $M_0=m_{\rho}$ and $f_0=f_{\rho}$ refer to the physical $\rho^0$ meson.

Note that the relation of the vector mesons $V_n$ conventionally used in
the GVMD model to physical $J^{P}=1^{-}$ vector mesons found in the Review of particle physics~\cite{Yao:2006px} is not direct.
The motivation for this is that while the vector meson masses 
are known with a reasonable accuracy up to $M_\rho \gtrsim 2$ GeV, 
there is no accurate data on the partial decay width
 $\Gamma_{e^+e^-}$ for mesons heavier than $\rho(1450)$. 
On the other hand, the parameterization~(\ref{eq:gvmd_2}) provides 
reasonable results for physical observables.
One can  check that the linear $n$-dependence of the ratio $M_n^2/M_0^2$ in Eq.~(\ref{eq:gvmd_2}) is confirmed experimentally for large-$n$~\cite{Yao:2006px}.
However, the slope of the $n$-dependence is underestimated
by approximately a factor of two.

It is important to point out that, at high energies, the DVCS cross section at the 
photon level is dominated by the contribution of the transversely polarized 
virtual photons due to the helicity conservation~\cite{Aktas:2005ty}.
 In the language of the color dipole model,
this dominance is explained by the dominance of the large transverse-size dipoles
 over the small transverse-size dipoles, see~\cite{Diehl:2003ny} for the discussion.
Therefore, Eq.~(\ref{eq:gvmd_1}) gives the complete description of the DVCS
amplitude.

The quantity $\Sigma_{n, n^{\prime}}(t)$ is the $V_{n,{\rm tr}} +p \to V_{n^{\prime},{\rm tr}}+p$
scattering amplitude, see Fig.~\ref{fig:dvcs_gvmd}. 
The matrix $\Sigma_{n, n^{\prime}}(t)$ is assumed to have a tri-diagonal form 
with the following non-zero elements,
\begin{eqnarray}
&&\Sigma_{n,n}(W,t)=i \sigma_{\rho p} (W^2)(1-i\eta) F_n(t) \,, \nonumber\\
&&\Sigma_{n,n+1}(W,t)=\Sigma_{n+1,n}(W,t)=
-\frac{1}{2} \frac{M_n}{M_{n+1}} \left(1-2\, \delta \frac{m^2_{\rho}}{M_n^2}\right)
\Sigma_{n,n}(W,t) \,,
\label{eq:gvmd_3}
\end{eqnarray}
where $\sigma_{\rho p}$ is the $\rho$ meson-proton cross section 
for the transversely polarized meson;
$\eta$ is the ratio of the real to imaginary parts of the $\rho$ meson-proton 
scattering amplitude; $\delta = 0.2$ is the parameter of the model.

The function $F_n(t)$ models the $t$-dependence of $\Sigma_{n, n^{\prime}}(t)$, 
which goes beyond the original formulation of the GVMD model~\cite{Fraas:1974gh,Ditsas:1975vd,Shaw:1993gx}, which addressed only
the forward $t=0$ limit. 
In our analysis, we use the following form of $F_n(t)$,
\begin{equation}
F_n(t)=\exp\left(-\frac{1}{2} \left[\frac{1}{n+1}B_{1}+\frac{n}{n+1}B_{2}\right]
 |t| \right) \,,
\label{eq:gvmd_4}
\end{equation}
where $B_1=11$ GeV$^{-2}$ and $B_{2}=4.3$ GeV$^{-2}$. The choice of the slopes
$B_1$ and $B_2$ is motivated as follows. 

For the moment, let us  
replace the final real photon by the $\rho$ meson. In the photoproduction limit,
the $\gamma p \to \rho p$ cross section measured at HERA by the H1 collaboration 
was fitted to the form $\exp(-B |t|)$
with the slope $B=(10.9 \pm 2.4 \pm 1.1)$ GeV$^{-2}$~\cite{Aid:1996bs}.
The ZEUS measurement gives essentially the same value of $B$~\cite{Breitweg:1997ed}.

In electroproduction, the slope of the exponential fit to the
$\gamma^{\ast} p \to \rho p$ cross section is much smaller than in photoproduction: It decreases from $B=(8.0 \pm 0.5 \pm 0.6)$ GeV$^{-2}$ at $Q^2=1.8$ GeV$^2$ to 
$B=(4.7 \pm 1.0 \pm 0.7)$ GeV$^{-2}$ at $Q^2=21.2$ GeV$^2$~\cite{Adloff:1999kg}. 

This decrease of the slope of the $t$-dependence with
increasing $Q^2$ is effectively parameterized by Eq.~(\ref{eq:gvmd_4})
as a decrease of the slope with the increasing number of the vector meson $n$.
Indeed, close to the photoproduction limit, the dominant contribution
to the sum in Eq.~(\ref{eq:gvmd_1})
comes from the $n=0$ term. In the opposite limit of large $Q^2$,
terms with large $n$, up to $M_n^2 \sim Q^2$, contribute to the sum.
Choosing $Q^2=0$ and 
$Q^2=21.2$ GeV$^2$ as reference points, we determine the values of the slopes
$B_1$ and $B_2$ in Eq.~(\ref{eq:gvmd_4}).

It is clear from the above discussion that while the value of the slope $B_1$ in
Eq.~(\ref{eq:gvmd_4}) is model-independent, the value of the slope
$B_2$ is somewhat more uncertain. 
We have chosen not to introduce $Q^2$-dependent slopes $B_1$ and $B_2$ since this
would contradict the spirit of the VMD model: The $W$ and $t$-dependence
of the DVCS amplitude is determined solely by the vector meson-proton scattering 
amplitudes; the vector meson propagators provide the $Q^2$-dependence.

It is important to note that the non-diagonal terms, $\Sigma_{n,n^{\prime}}$
with $n \neq n^{\prime}$, are essential in the GVMD model: The infinite series
in  Eq.~(\ref{eq:gvmd_1}) would have been divergent without the non-diagonal transitions. Also, the non-diagonal terms provide the correct scaling of
the total $\gamma^{\ast}p$ cross section.

Using Eqs.~(\ref{eq:gvmd_2}) and (\ref{eq:gvmd_3}), the DVCS amplitude in
Eq.~(\ref{eq:gvmd_1}) can be written in the following form 
\begin{eqnarray}
{\cal A}(\gamma_{{\rm tr}}^{\ast}+p \to \gamma+ p)&=&
i \frac{2(1+\delta)e^2}{f_{\rho}^2} \sigma_{\rho p} (W^2)(1-i\eta)
\sum_{n=0}^{\infty} 
\frac{F_n(t)}{(\frac{Q^2}{m_{\rho}^2}+1+2n)(\frac{Q^2}{m_{\rho}^2}+3+2n)}
\nonumber\\
&\times&\left[1+\frac{Q^2}{2 m_{\rho}^2 (1+\delta)(3 +2n)}\left(1+4 \delta \frac{1+n}{(1+2n)} \right)\right]  \,.
\label{eq:gvmd_5}
\end{eqnarray} 
Equation~(\ref{eq:gvmd_5}) involves four quantities, $f_{\rho}$, $\sigma_{\rho p}$,
$\eta$ and $\delta$, which are known with a certain degree of uncertainty.
One can reduce this uncertainty by expressing the DVCS amplitude in terms
of the total $\gamma p$ cross section,
\begin{eqnarray}
\sigma_{{\rm tot}}^{\gamma p}(W^2)&=&\Im m \, {\cal A}(\gamma+p \to \gamma+ p)_{|t=0}=
\frac{2(1+\delta)e^2}{f_{\rho}^2} \sigma_{\rho p} (W^2)
\sum_{n=0}^{\infty} \frac{1}{(1+2n)(3+2n)} \nonumber\\
&=&\frac{(1+\delta)e^2}{f_{\rho}^2} \sigma_{\rho p} (W^2) \,.
\label{eq:gvmd_6}
\end{eqnarray}
Therefore, the final expression for the DVCS amplitude reads
\begin{eqnarray}
{\cal A}(\gamma_{{\rm tr}}^{\ast}+p \to \gamma+ p)&=&
i 2\,\sigma_{{\rm tot}}^{\gamma p} (W^2)(1-i\eta)
\sum_{n=0}^{\infty} 
\frac{F_n(t)}{(\frac{Q^2}{m_{\rho}^2}+1+2n)(\frac{Q^2}{m_{\rho}^2}+3+2n)}
\nonumber\\
&\times&\left[1+\frac{Q^2}{2 m_{\rho}^2 (1+\delta)(3 +2n)}\left(1+4 \delta \frac{1+n}{(1+2n)} \right)\right]  \,.
\label{eq:gvmd_7}
\end{eqnarray} 
One should also note that another advantage of expressing the DVCS amplitude
in terms of $\sigma_{{\rm tot}}^{\gamma p}$ is that Eq.~(\ref{eq:gvmd_7})
effectively takes into account the contributions of the $\omega$ and 
$\phi$ vector mesons, which enter through the 
phenomenological parameterization of  $\sigma_{{\rm tot}}^{\gamma p}$.

In our analysis, we use the ZEUS parameterization of 
$\sigma_{{\rm tot}}^{\gamma p}(W^2)$~\cite{Chekanov:2001gw}
\begin{equation}
\sigma_{{\rm tot}}^{\gamma p}(W^2)=57\,W^{0.2}+121\,W^{-0.716} \,,
 \label{eq:gvmd_8}
\end{equation}  
where the cross section is in $\mu$b and $W$ is in GeV.

The ratio of the real to imaginary parts of the
 $V +p \to V+p$
scattering amplitude, $\eta$, is found using the Gribov-Migdal relation~\cite{Gribov:1968uy}, 
\begin {equation}
\eta \approx \frac{\pi}{2}\, \frac{p}{2} \approx 0.16 \,,
\label{eq:Gribov-Migdal}
\end{equation}
where $p=0.2$ was used, which corresponds to the power of the 
$W$-dependence of $\sigma_{{\rm tot}}^{\gamma p}(W^2)$
at large $W$ in Eq.~(\ref{eq:gvmd_8}).

The remaining parameter in Eq.~(\ref{eq:gvmd_7}) is $\delta$, for which
we use $\delta=0.2$~\cite{Fraas:1974gh,Ditsas:1975vd,Shaw:1993gx,Pautz:1997eh}.
However, the exact numerical value of $\delta$ affects weakly 
our numerical predictions.

One of simplest DVCS observables is the skewing factor $R$, which is 
defined as the ratio 
of the DVCS to the DIS amplitudes and which was recently extracted from the HERA
DVCS and DIS data~\cite{Schoeffel:2007dt},
\begin{equation}
R(t) \equiv \frac{\Im m {\cal A}(\gamma^{\ast}_{{\rm tr}}+p \to \gamma+p)_{|t}}
{\Im m {\cal A}(\gamma^{\ast}_{{\rm tr}}+p \to \gamma^{\ast}_{{\rm tr}}+p)_{|t=0}} \,.
\label{eq:R_1}
\end{equation}
Note that we generalized the ratio $R$ originally defined at $t=t_{{\rm min}}$~\cite{Schoeffel:2007dt}
 to any value of $t$. At high energies, the minimal
momentum transfer $|t_{{\rm min}}| \approx x_B^2 m_N^2 \approx 0$,
where $x_B$ is the Bjorken variable; $m_N$ is the nucleon mass.

The GVMD model makes an unambiguous prediction for the ratio $R$,
\begin{eqnarray}
R(t)&=& 
\sum_{n=0}^{\infty} \frac{F_n(t)}{(\frac{Q^2}{m_{\rho}^2}+1+2n)(\frac{Q^2}{m_{\rho}^2}+3+2n)}
\left[1+\delta+\frac{Q^2}{2 m_{\rho}^2(3 +2n)}\left(1+4 \delta \frac{1+n}{(1+2n)} \right)\right]
\nonumber\\
&\Big/&\left(\sum_{n=0}^{\infty} \frac{1+2n}{(\frac{Q^2}{m_{\rho}^2}+1+2n)^2
(\frac{Q^2}{m_{\rho}^2}+3+2n)}+\frac{1}{2} \frac{\delta}{1+\frac{Q^2}{m_{\rho}^2}}
\right)
 \,.
\label{eq:R_2}
\end{eqnarray}

\begin{figure}[t]
\begin{center}
\epsfig{file=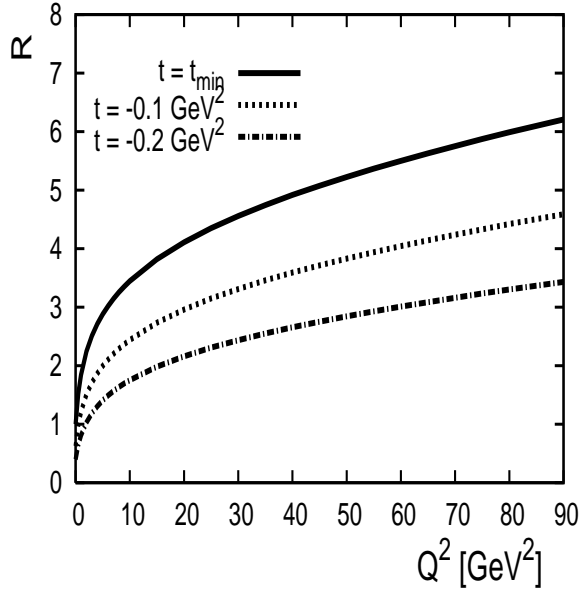,width=8cm,height=8cm}
\caption{The GVMD prediction for the
ratio $R(t)$ of the DVCS and DIS amplitudes, see Eq.~(\ref{eq:R_1}),
as a function of $Q^2$ for three values of $t$.}
\label{fig:Schoeffel}
\end{center}
\end{figure}

Figure~\ref{fig:Schoeffel} presents
the GVMD predictions for the ratio $R$ as a function
of $Q^2$ for three values of $t$: $t=t_{{\rm min}} \approx 0$ (evaluated with $x_B=0.001$), $t=-0.1$ GeV$^2$ and 
$t=-0.2$ GeV$^2$.
 Note that in the GVMD model, the ratio $R$ 
does not depend on $W$ or $x_B$ at given $Q^2$ and $t$.

A comparison of the solid curve in Fig.~\ref{fig:Schoeffel} to the 
experimental results for the ratio $R$, see Fig.~4 of Ref.~\cite{Schoeffel:2007dt},
reveals that the GVMD model provides a good description of the data for 
$Q^2 < 5$ GeV$^2$. For higher values of $Q^2$, the GVMD model overestimates the
experimental $R$. Therefore, the GVMD model and similar models  can be used 
to reliably determine DVCS observables and generalized parton distributions at $Q^2$ of
the order of a few GeV$^2$. This can be used as an input for QCD evolution
to higher $Q^2$ scales. An example of such an approach, 
which uses the align-jet model to construct input GPDs and
which excellently compares to 
the HERA data on the DVCS cross section and on the ratio $R$, was worked 
out in~\cite{Freund:2002qf}.

\subsection{DVCS cross section}

The DVCS amplitude in Eq.~(\ref{eq:gvmd_7}) is normalized such that in the 
$Q^2 \to 0$ limit, the imaginary part of the $\gamma p \to \gamma p$ 
amplitude
is equal to the
total photoabsorption cross section, see Eq.~(\ref{eq:gvmd_6}). With such a normalization,
the differential and integrated DVCS cross sections at the photon level read
\begin{eqnarray}
&&\frac{d \sigma_{{\rm DVCS}}}{dt}(W,Q^2,t)=\frac{1}{16 \pi} |{\cal A}(\gamma_{{\rm tr}}^{\ast}+p \to \gamma+ p)|^2 \,, \nonumber\\
&&\sigma_{{\rm DVCS}}(W,Q^2)=\frac{1}{16 \pi} \int_{-1\ {\rm GeV}^2}^{t_{\rm min}}
dt \,|{\cal A}(\gamma_{{\rm tr}}^{\ast}+p \to \gamma+ p)|^2 \,.
\label{eq:cs_1}
\end{eqnarray}
where $t_{{\rm min}} \approx -x_B^2 m_N^2$
($t_{\rm min} \approx 0$ in the HERA kinematics).

In order to compare the GVMD model predictions to the  data on the DVCS cross 
section at the photon level~\cite{Chekanov:2003ya,Aktas:2005ty},
 one needs to make sure that one compares the same 
quantities. Using the classic result of L.~N.~Hand~\cite{Hand:1963bb},
one can readily see that the HERA DVCS cross section at the photon level
is indeed a properly defined and normalized cross section of the
$\gamma^{\ast}p \to \gamma p$ reaction.

As a byproduct of the above mentioned exercise, one establishes the connection
between the GVMD and GPD descriptions of the DVCS cross section:
\begin{eqnarray}
|{\cal A}(\gamma_{{\rm tr}}^{\ast}+p \to \gamma+ p)|^2&=&\frac{e^4 x_B^2}{Q^4 \sqrt{1+\epsilon^2}} 
\Big((1-\xi^2)
(|{\cal H}|^2+ |\tilde{{\cal H}}|^2) 
-\xi^2 ({\cal H}^{\ast} {\cal E}+{\cal H} {\cal E}^{\ast}
\nonumber\\
 &+& \tilde{{\cal H}^{\ast}}\tilde{{\cal E}}+{\tilde {\cal H}} \tilde{{\cal E}^{\ast}})-
|{\cal E}|^2 (\frac{t}{4 m_N^2}+\xi^2)-\xi^2 \frac{t}{4 m_N^2} |\tilde{{\cal E}}|^2
 \Big) \,,
\label{eq:cs_2}
\end{eqnarray}
where $x_B$ is the Bjorken variable; $\xi=x_B/(2-x_B)$;
$\epsilon^2=4 x_B^2 m_N^2/Q^2$.
The quantities ${\cal H}$, ${\cal E}$, $\tilde{{\cal H}}$ and ${\tilde {\cal E}}$
are the so-called Compton form factors of the corresponding proton GPDs~\cite{Belitsky:2001ns}.
It is important to have the connection between the GVMD-based  and the GPD-based
descriptions of the DVCS cross section since the both approaches have an
overlapping region of applicability, namely, $1 < Q^2 < 5$ GeV$^2$.

The simple expression for the DVCS amplitude in the GVMD model~(\ref{eq:gvmd_7})
allows one to examine the relative contribution of $1/Q^2$-corrections, which correspond to higher-twist corrections in perturbative QCD. To this end, let us 
expand the DVCS amplitude in Eq.~(\ref{eq:gvmd_7}) in terms of $1/Q^2$ and let us call
the leading contribution, which behaves as $1/Q^2$, 
${\cal A}^{{\rm LO}}(\gamma_{{\rm tr}}^{\ast}+p \to \gamma+ p)$. The corresponding
$t$-integrated cross section is denoted as $\sigma_{{\rm DVCS}}^{{\rm LO}}$.

We quantify the contribution of $1/Q^2$-corrections to the DVCS amplitude and to
the DVCS cross section by introducting the ratios
 $R_{{\rm ampl}}^{{\rm HT}}$ and  $R_{\sigma}^{{\rm HT}}$,
\begin{eqnarray}
R_{{\rm ampl}}^{{\rm HT}}(Q^2)&=&1-\frac{{\cal A}^{{\rm LO}}(\gamma_{{\rm tr}}^{\ast}+p \to \gamma+ p)_{|t=t_{{\rm min}}}}{{\cal A}(\gamma_{{\rm tr}}^{\ast}+p \to \gamma+ p)_{|t=t_{{\rm min}}}} \,, \nonumber\\
R_{\sigma}^{{\rm HT}}(Q^2)&=&1-\frac{\sigma_{{\rm DVCS}}^{{\rm LO}}}{\sigma_{{\rm DVCS}}} \,.
\label{eq:HT}
\end{eqnarray}
The ratios $R_{{\rm ampl}}^{{\rm HT}}$ and $R_{\sigma}^{{\rm HT}}$
as functions of $Q^2$ are summarized in Table~\ref{table:HT}.
Note that these ratios do not depend on $W$ in the chosen model.
\begin{table}
\caption{The $1/Q^2$-corrections to the DVCS amplitude and to the 
$t$-integrated DVCS cross section as functions of $Q^2$, see Eq.~(\ref{eq:HT}).}
\label{table:HT}
\begin{ruledtabular}
\begin{tabular}{lll}
$Q^2$ [GeV$^2$] & $R_{{\rm ampl}}^{{\rm HT}}(Q^2)$ & $R_{\sigma}^{{\rm HT}}(Q^2)$ \\
\hline
2 & 0.20 & 0.56 \\
4 & 0.11 & 0.32 \\
8 & 0.058 & 0.17 \\
\end{tabular}
\end{ruledtabular}
\end{table}

As one can see from Table~\ref{table:HT}, the $1/Q^2$-corrections 
are large. Moreover, $R_{\sigma}^{{\rm HT}}(Q^2) > 2 R_{{\rm ampl}}^{{\rm HT}}(Q^2)$ due to the enhancement of the heavy vector meson contributions
to $\sigma_{{\rm DVCS}}$ because of the decreasing slope of the 
$t$-dependence of the DVCS amplitude with increasing $n$, 
see Eq.~(\ref{eq:gvmd_4}).

\subsection{Comparison to the HERA DVCS data}

Using Eqs.~(\ref{eq:gvmd_7}) and (\ref{eq:cs_1}), we make predictions for 
the DVCS cross section and compare our findings to the HERA data~\cite{Chekanov:2003ya,Aktas:2005ty}.    

\begin{figure}[t]
\begin{center}
\epsfig{file=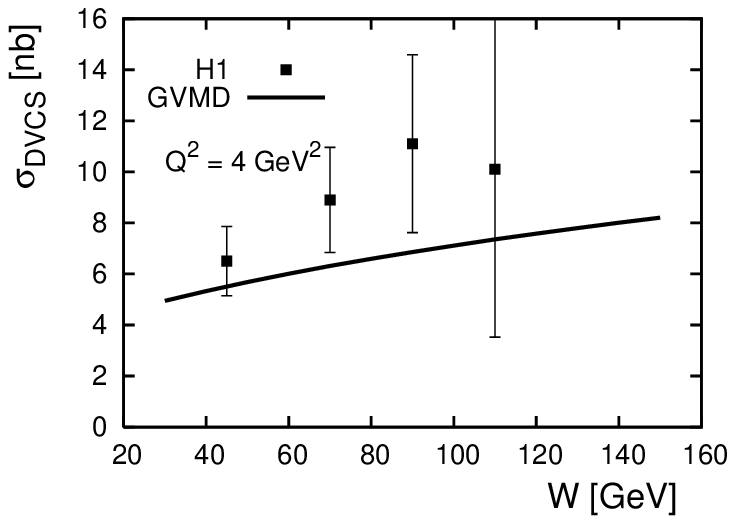,width=8cm,height=8cm}
\epsfig{file=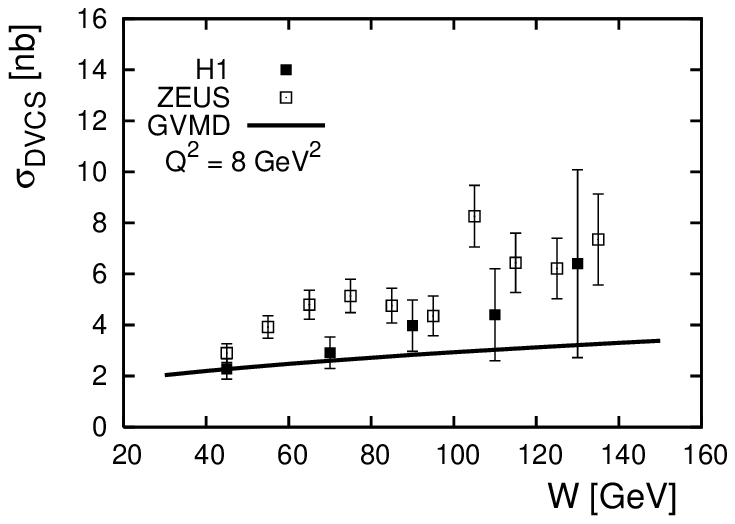,width=8cm,height=8cm}
\caption{The DVCS cross section as a function of $W$.
The GVMD model results (solid curves) are compared to the H1~\cite{Aktas:2005ty} and ZEUS~\cite{Chekanov:2003ya} data.
The error bars correspond to the statistical and systematic errors added in quadrature.
}
\label{fig:gvmd_wdep}
\end{center}
\end{figure}

Figure~\ref{fig:gvmd_wdep} presents the $W$-dependence of the DVCS cross
section at $Q^2=4$ GeV$^2$ and $Q^2=8$ GeV$^2$.
The solid curves correspond to the GVMD calculations; the experimental points are those
from the H1~\cite{Aktas:2005ty} and ZEUS~\cite{Chekanov:2003ya} measurements.
The error bars correspond to the statistical and systematic errors added in quadrature.
The ZEUS data taken at $Q^2=9.6$ GeV$^2$ have been interpolated to $Q^2=8.0$ GeV$^2$
using the fit to the $Q^2$-dependence of $\sigma_{{\rm DVCS}}$,
 $\sigma_{{\rm DVCS}} \sim 1/(Q^2)^n$ with
$n=1.54$~\cite{Chekanov:2003ya}.

We shall discuss the left and right panels of Fig.~\ref{fig:gvmd_wdep}
separately. 
As seen from the left panel of Fig.~\ref{fig:gvmd_wdep},
the GVMD model reproduces both the absolute value and the 
$W$-dependence of $\sigma_{{\rm DVCS}}$  sufficiently  well. The latter fact signifies that,
at $Q^2=4$ GeV$^2$, the DVCS cross section is still dominated by soft physics.
At $Q^2=4$ GeV$^2$, the $W$-behavior  of $\sigma_{{\rm DVCS}}$ is consistent
with that predicted by the GVMD model,
$\sigma_{{\rm DVCS}} \sim W^{0.4}$.

Turning to the right panel of Fig.~\ref{fig:gvmd_wdep}, we observe that 
while the GVMD model compares fairly with the H1 data, the model 
underestimates the slope of the $W$-dependence of $\sigma_{{\rm DVCS}}$ 
for the ZEUS data set, which has smaller error bars.
In particular, the predicted $\sigma_{{\rm DVCS}} \sim W^{0.4}$ behavior is much slower than that given by the fit to the ZEUS data points,
$\sigma_{{\rm DVCS}} \sim W^{\delta}$ with $\delta=0.75 \pm 0.15$~\cite{Chekanov:2003ya}.
This indicates the onset of the hard regime in the total DVCS cross section at $Q^2=8$ GeV$^2$, where the GVMD model becomes inadequate.

\begin{figure}[t]
\begin{center}
\epsfig{file=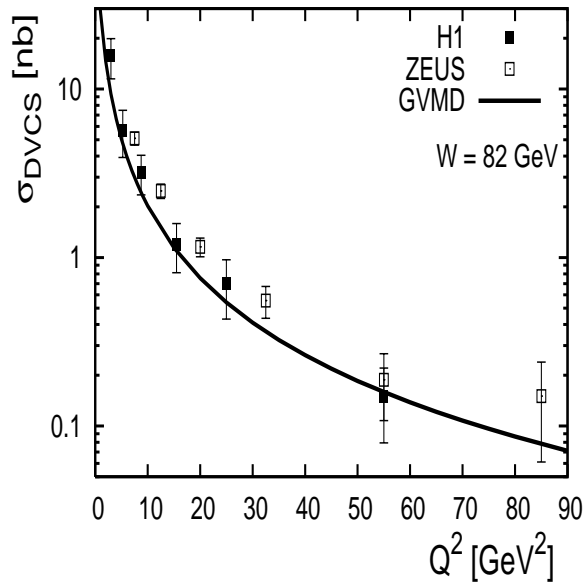,width=8cm,height=8cm}
\caption{The DVCS cross section as a function of $Q^2$.
The GVMD model result (solid curve) is compared to the H1~\cite{Aktas:2005ty} and ZEUS~\cite{Chekanov:2003ya} data.
The error bars correspond to the statistical and systematic errors added in quadrature.
}
\label{fig:gvmd_q2dep}
\end{center}
\end{figure}

Figure~\ref{fig:gvmd_q2dep} presents the $Q^2$-dependence of the DVCS cross
section at $W=82$ GeV.
The solid curve corresponds to the GVMD model; the experimental points 
come from the H1~\cite{Aktas:2005ty} and ZEUS~\cite{Chekanov:2003ya} 
experiments.
The error bars correspond to the statistical and systematic errors added in quadrature.
The ZEUS data taken at $W=89$ GeV have been extrapolated to $W=82$ GeV
using the fitted $W$-dependence of $\sigma_{{\rm DVCS}}$,
 $\sigma_{{\rm DVCS}} \sim W^{0.75}$~\cite{Chekanov:2003ya}.

One sees from Fig.~\ref{fig:gvmd_q2dep} that the GVMD model reproduces the
$Q^2$-dependence of $\sigma_{{\rm DVCS}}$ over a very wide range of
$Q^2$, $3 \leq Q^2 \leq 85$ GeV$^2$. This is quite a remarkable result that
the model, which was initially developed for photoproduction and was 
later extended to
electroproduction with $Q^2$ of the order of a few GeV$^2$,
provides a quantitative description for such large values of 
$Q^2$. 
In other words, at fixed $W$, the GVMD model correctly reproduces the
$Q^2$-scaling of $\sigma_{{\rm DVCS}}$.

\begin{figure}[t]
\begin{center}
\epsfig{file=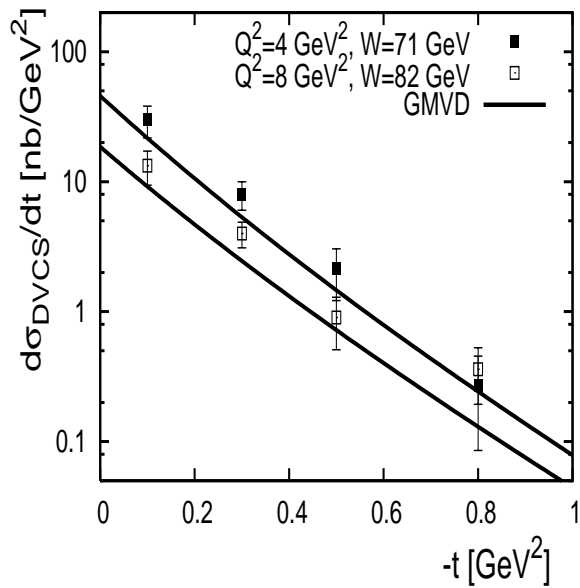,width=8cm,height=8cm}
\caption{The differential DVCS cross section $d \sigma_{{\rm DVCS}}/dt$
 as a function of $t$.
The GVMD model calculation (solid curves) is compared to the 
H1 data~\cite{Aktas:2005ty}.
The error bars are the statistical and systematic errors added in quadrature.
}
\label{fig:gvmd_tdep}
\end{center}
\end{figure}

Figure~\ref{fig:gvmd_tdep} presents the comparison
of the $t$-dependence
 of the GVMD model
calculations (solid curves) to the H1 data on the differential DVCS
cross section $d \sigma_{{\rm DVCS}}/dt$~\cite{Aktas:2005ty}.
The error bars are the statistical and systematic errors added in quadrature.

As one sees from Fig.~\ref{fig:gvmd_tdep}, the GVMD model describes 
$d \sigma_{{\rm DVCS}}/dt$  well. This result is not quite trivial.
In order to achieve this within the framework of the GVMD model, one has 
to assume that either all vector meson $V_n$-nucleon cross sections 
have the same $Q^2$-dependent slope of the $t$-dependence 
or that the slope 
decreases with increasing $n$, see Eq.~(\ref{eq:gvmd_4}).
While the value of the slope
$B_1=11$ GeV$^{-2}$ is fixed by photoproduction of $\rho$ mesons,
the value of the slope $B_2$ is model-dependent. The values $B_2=4 \div 5$
GeV$^{-2}$ provide a good agreement with the H1 data 
(see Fig.~\ref{fig:gvmd_tdep}), which were fitted to the exponential form,
$d \sigma_{{\rm DVCS}}/dt \sim e^{-B |t|}$ with $B=(6.66 \pm 0.54 \pm 0.43)$
 GeV$^{-2}$ at $Q^2=4$ GeV$^2$ and 
$B=(5.82 \pm 0.59 \pm 0.60)$
 GeV$^{-2}$ at $Q^2=8$ GeV$^2$~\cite{Aktas:2005ty}.

\subsection{DVCS cross section in Jefferson Lab kinematics}

The $\vec{e}p \to e p \gamma$ cross section in the DVCS regime was
recently measured by the Hall~A collaboration at Jefferson 
Laboratory (JLab)~\cite{Munoz Camacho:2006hx}. The cross section involves the contributions
of the Bethe-Heitler (BH) amplitude squared, the DVCS amplitude squared and 
the interference of the BH and DVCS amplitudes. 
Based on the kinematics of the experiment,
in the analysis of the data,
the contribution of the DVCS amplitude squared was neglected compared to the other
two contributions~\cite{Belitsky:2001ns}, 
which allowed for the extraction of the so-called Compton 
form factors of the proton.

In this subsection, we check the validity of the assumption that 
the contribution of the DVCS amplitude squared is negligibly small by explicitly
calculating the DVCS cross section within the GVMD model 
in the Jefferson Lab kinematics.
The DVCS cross section at the lepton level reads, see e.g.~\cite{Belitsky:2001ns},
\begin{equation}
\frac{d^4 \sigma}{dQ^2 dx_B dt d \phi}=\frac{\alpha_{{\rm e.m.}} (1-y+y^2/2)}{\pi Q^2 x_B} \frac{1}{2 \pi} \frac{d \sigma_{{\rm DVCS}}(W,Q^2,t)}{dt} \,,
\label{eq:sigma_jlab}
\end{equation}
where $\alpha_{{\rm e.m.}}$ is the fine-structure constant;
$\phi$ is the angle between the lepton and production planes;
 $d \sigma_{{\rm DVCS}}(W,Q^2,t)/dt$ is the DVCS cross section at the proton
level defined by Eq.~(\ref{eq:cs_1}). The extra factor $1/(2\pi)$ in the right-hand 
side of Eq.~(\ref{eq:sigma_jlab}) takes into account the fact that the integration over 
the angle $\phi$ is included in the definition of $d \sigma_{{\rm DVCS}}(W,Q^2,t)/dt$.
The DVCS cross section does not depend on 
$\phi$ when one neglects the photon helicity changing transitions~\cite{Diehl:2000xz}.

Using Eqs.~(\ref{eq:gvmd_7}) and (\ref{eq:cs_1}), we evaluate 
the DVCS cross section at the lepton level
$d^4 \sigma/(dQ^2 dx_B dt d \phi)$ in the kinematics of the Hall A experiment,
$E=5.75$ GeV (the energy of the lepton beam), $Q^2=2.3$ GeV, $t=-0.28$ GeV and
$x_B=0.36$,
\begin{equation}
\frac{d^4 \sigma}{dQ^2 dx_B dt d \phi}=0.0022 \ {\rm nb}/{\rm GeV}^4 \,.
\end{equation}
This value is an order of magnitude smaller than the sum of the BH and interference
contributions to the $ep \to e p \gamma$ cross section, which confirms the assumption
that, in the JLab kinematics, the contribution of the DVCS amplitude squared to the
$ep \to e p \gamma$ cross section can be safely neglected.

\section{DVCS on nuclei}
\label{sec:dvcs-nucleus}

In this section, we derive the expression for the  DVCS
amplitude on a nucleus using the GVMD model for the photon-nucleon interactions
and the generalized Glauber 
formalism~\cite{Bauer:1977iq}
in order to account for the multiple rescattering of the vector mesons 
inside the nucleus. Using the obtained amplitude, we make predictions for the
nuclear DVCS cross section at the photon level in the collider kinematics.


At high energies,
in the GVMD model, photons (real and virtual) interact with hadrons
by fluctuating into an infinite sum of vector mesons.
When the involved hadron is a nucleus,
 each vector meson undergoes multiple interactions
with the nucleons of the nucleus, which leads to the attenuation (decrease) of
the vector meson-nucleus cross section compared to the sum of free vector
meson-nucleon cross section. As a consequence, the resulting photon-nucleus
cross section is smaller than the sum of the corresponding
photon-nucleon cross sections. This phenomenon is called nuclear shadowing.
It has been observed in various reactions with nuclei induced by real and
virtual photons, see~\cite{Bauer:1977iq,Piller:1999wx} for review.

\begin{figure}[t]
\begin{center}
\epsfig{file=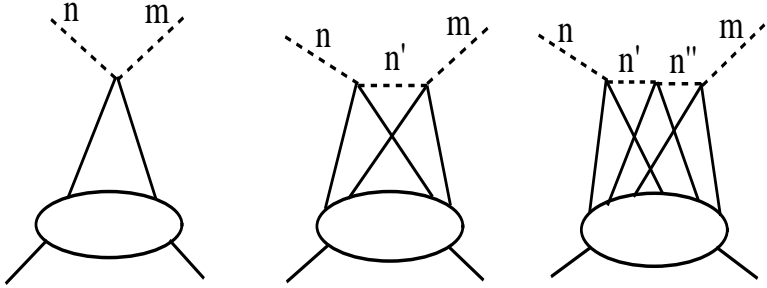,width=12cm,height=7cm}
\caption{A schematic representation of the multiple scattering (Glauber)
series for the nuclear DVCS amplitude in the GVMD model.
The dashed lines correspond to vector mesons; the solid lines correspond to nucleons;
the ovals with legs correspond to the final and initial nucleus.}
\label{fig:nucl_dvcs_gvmd}
\end{center}
\end{figure}

In the GVMD model, the nuclear DVCS amplitude can 
be organized as a multiple scattering series (Glauber series), 
where each term corresponds to the number 
of interactions of the vector mesons with the nucleons. This is schematically
presented in Fig.~\ref{fig:nucl_dvcs_gvmd}, where the interactions with one, two and three nucleons are depicted. The dashed lines correspond to the vector mesons
(note that the GVMD model allows for non-diagonal vector meson-nucleon
transitions);
the solid lines correspond to the nucleons involved in the interactions; the nuclear
part is denoted by ovals with legs corresponding to the initial and final nucleus.

Using the standard technique~\cite{Bauer:1977iq},
one can readily write down the expression for the nuclear DVCS amplitude
${\cal A}(\gamma^{\ast}+A \to \gamma+A)$,
\begin{eqnarray}
{\cal A}(\gamma^{\ast}+A \to \gamma+A)&=&\sum_{n,m=0}^{\infty} \frac{e}{f_n}
\frac{M_n^2}{M_n^2+Q^2}  \frac{e}{f_m} \Bigg[A F_A(t) \Sigma_{n,m}(W,t) \nonumber\\
&-&\frac{A(A-1)}{2 i} \int_{-\infty}^{\infty} dz_1
\int_{z_1}^{\infty} dz_2 \int d^2 \vec{b} \,e^{i \vec{q}_t \cdot \vec{b}}\rho(b,z_1) 
\rho(b,z_2) e^{iz_1(k_{\gamma^{\ast}}-k_{V_n})}
\nonumber\\
&\times &\tilde{\Sigma}_{n,n^{\prime}}(W,z_1) 
\Big(\delta_{n^{\prime},n^{\prime \prime}}-\frac{A-2}{2i}\,\tilde{\Sigma}_{n^{\prime},n^{\prime \prime}}(W,z^{\prime})
\Theta(z_1 \leq z^{\prime} \leq z_2) \rho(b,z^{\prime}) \nonumber\\
&+& \dots \Big) 
\tilde{\Sigma}_{n^{\prime \prime},m}(W,z_2) e^{iz_2(k_{V_m}-k_{\gamma})}
\Bigg] \,,
\label{eq:nucl1}
\end{eqnarray}
where $A$ is the number of nucleons in the nucleus (we do not distinguish
protons and neutrons);
$F_A(t)$ is the nuclear form factor ($F_A(0)=1$);
$\rho(r)$ is the density of nucleons in the nucleus~\cite{DeJager:1987qc}; $\vec{b}$ is the two-dimensional
vector (impact parameter) in the plane perpendicular to the direction of 
the incoming photon,
whose momentum is assumed to be along the $z$-direction;
$z_i$ are longitudinal positions of the nucleons of the nucleus involved in the 
interaction;
$\vec{q}_t$ is the transverse component of the momentum transfer.
 Since we neglected the $t$-dependence of the elementary vector meson-nucleon
scattering amplitudes compared to the steep $t$-dependence of the nuclear form factor,
all scatterings of the vector mesons in Eq.~(\ref{eq:nucl1}) occur
at the same impact parameter $\vec{b}$.

In Eq.~(\ref{eq:nucl1}), 
\begin{equation}
\tilde{\Sigma}_{n,m}(W,z)=e^{i z(k_{V_n}-k_{V_m})}\Sigma_{n,m}(W,t=0) \,,
\label{eq:nucl2}
\end{equation}
where $\Sigma_{n,m}$ is defined by Eq.~(\ref{eq:gvmd_3}).
In Eqs.~(\ref{eq:nucl1}) and (\ref{eq:nucl2}), the exponential factors
(except for the $\exp(i \vec{q}_t \cdot \vec{b})$ factor) arise due to the
non-zero longitudinal momentum transfer associated with 
non-diagonal in mass transitions. At high energies,
\begin{eqnarray}
&&k_{\gamma^{\ast}}-k_{V_n}=\sqrt{\nu^2+Q^2}-\sqrt{\nu^2-M_n^2}=\frac{Q^2+M_n^2}{2 \nu}=
x_B m_N \left(1+\frac{M_n^2}{Q^2} \right) \,, \nonumber\\
&&k_{V_m}-k_{\gamma}=\sqrt{\nu^{\prime 2}-M_n^2}-\nu^{\prime}=-\frac{M_m^2}{2 \nu^{\prime}}
\approx -\frac{M_m^2}{2 \nu}=-x_B m_N \frac{M_m^2}{Q^2} \,,\nonumber\\
&&k_{V_n}-k_{V_m}=\frac{M_m^2-M_n^2}{2 \nu} \,,
\label{eq:nucl3}
\end{eqnarray}
where $\nu$ is the energy of the incoming virtual photon in the laboratory reference
frame; $\nu^{\prime}$ is the energy of the final real photon. We also used that
$\nu^{\prime}=\nu+t/(2 m_N) \approx \nu$ for the small momentum transfer $t$.

In Eq.~(\ref{eq:nucl1}), the first term corresponds to the left graph in 
Fig.~\ref{fig:nucl_dvcs_gvmd}, which describes the interaction with one nucleon of
the nucleus (the Born term). The second term in Eq.~(\ref{eq:nucl1}) 
corresponds to the middle
graph in Fig.~\ref{fig:nucl_dvcs_gvmd}, which describes the interaction of hadronic 
fluctuations of the involved photons with two nucleons of the nucleus.
Those nucleons are located at the points $\vec{r}_1=(\vec{b},z_1)$ and
$\vec{r}_2=(\vec{b},z_2)$. 
This graph leads to the attenuation (nuclear shadowing) of the Born term.
The third term corresponds to the interaction with three nucleons of the nucleus.
The dots in Eq.~(\ref{eq:nucl1}) denote higher rescattering terms not shown in
Fig.~\ref{fig:nucl_dvcs_gvmd}.

Equation~(\ref{eq:nucl1}) is rather general and, because of the non-diagonal
$V_n \to V_{n^{\prime}}$ transitions, the direct calculation of the nuclear
DVCS amplitude for heavy nuclei using Eq.~(\ref{eq:nucl1}) is impossible.
Therefore, for our numerical predictions,
we make an approximation and
 ignore the non-diagonal transitions for the
interactions with three and more nucleons (this does affect the convergence of the
series),
\begin{eqnarray}
&&\delta_{n^{\prime},n^{\prime \prime}}-\frac{A-2}{2i}\,\tilde{\Sigma}_{n^{\prime},n^{\prime \prime}}(W,z^{\prime})
\Theta(z_1 \leq z^{\prime} \leq z_2) \rho(b,z^{\prime}) 
+ \dots  \nonumber\\
&\to& \delta_{n^{\prime},n^{\prime\prime}}
\left(1-\frac{A-2}{2i}\,\Sigma_{n^{\prime},n^{\prime}}(W,t=0)
\Theta(z_1 \leq z^{\prime} \leq z_2) \rho(b,z^{\prime})
+ \dots \right) \nonumber\\
&=&\delta_{n^{\prime},n^{\prime\prime}}\, e^{-\frac{A}{2} \sigma_{\rho p} (W^2)(1-i\eta) \int^{z_2}_{z_1} dz^{\prime} \rho(b,z^{\prime})} \,,
\label{eq:nucl4}
\end{eqnarray}
where in the last line we used the large-$A$ approximation.
Therefore, Eq.~(\ref{eq:nucl1}) now reads
\begin{eqnarray}
{\cal A}(\gamma^{\ast}+A &\to& \gamma+A)=\sum_{n,m=0}^{\infty} \frac{e}{f_n}
\frac{M_n^2}{M_n^2+Q^2}  \frac{e}{f_m} \Bigg[A F_A(t) \Sigma_{n,m}(W,t) \nonumber\\
&-&\frac{A(A-1)}{2 i} \int_{-\infty}^{\infty} dz_1
\int_{z_1}^{\infty} dz_2 \int d^2 \vec{b} \,e^{i \vec{q}_t \cdot \vec{b}}\rho(b,z_1) 
\rho(b,z_2) e^{iz_1(k_{\gamma^{\ast}}-k_{V_n})}
\nonumber\\
&\times &\tilde{\Sigma}_{n,n^{\prime}}(W,z_1) 
 e^{-\frac{A}{2} \sigma_{\rho p} (W^2)(1-i\eta) \int^{z_2}_{z_1} dz^{\prime}
\rho(b,z^{\prime})}
\tilde{\Sigma}_{n^{\prime},m}(W,z_2) e^{iz_2(k_{V_m}-k_{\gamma})}
\Bigg] \,.
\label{eq:nucl5}
\end{eqnarray}

For comparison of nuclear shadowing in DVCS and DIS, we also give the expression 
for the forward nuclear DIS amplitude, which can be readily obtained from 
Eq.~(\ref{eq:nucl5}), 
\begin{eqnarray}
{\cal A}(\gamma^{\ast}+A &\to& \gamma^{\ast}+A)_{|t=0}=\sum_{n,m=0}^{\infty} \frac{e}{f_n}
\frac{M_n^2}{M_n^2+Q^2}  \frac{e}{f_m} \frac{M_m^2}{M_m^2+Q^2}\Bigg[A \Sigma_{n,m}(W,t=0) \nonumber\\
&-&\frac{A(A-1)}{2 i} \int_{-\infty}^{\infty} dz_1
\int_{z_1}^{\infty} dz_2 \int d^2 \vec{b} \,\rho(b,z_1) 
\rho(b,z_2) e^{iz_1(k_{\gamma^{\ast}}-k_{V_n})}
\nonumber\\
&\times &\tilde{\Sigma}_{n,n^{\prime}}(W,z_1) 
 e^{-\frac{A}{2} \sigma_{\rho p} (W^2)(1-i\eta) \int^{z_2}_{z_1} dz^{\prime}
\rho(b,z^{\prime})}
\tilde{\Sigma}_{n^{\prime},m}(W,z_2) e^{iz_2(k_{V_m}-k_{\gamma^{\ast}})}
\Bigg] \,.
\label{eq:nucl5b}
\end{eqnarray}

We quantify predictions of the GVMD model for the nuclear DVCS and DIS
amplitudes by considering the ratios $R_{{\rm ampl}}^{{\rm Im}}$,
$R_{{\rm ampl}}^{{\rm Re}}$
 and $R_{{\rm ampl}}^{{\rm DIS}}$,
\begin{eqnarray}
R_{{\rm ampl}}^{{\rm Im}} & = & \frac{\Im m\,{\cal A}(\gamma^{\ast}+A \to \gamma+A)}
{\Im m \,{\cal A}^{{\rm Born}}(\gamma^{\ast}+A \to \gamma+A)} \,, \nonumber\\
R_{{\rm ampl}}^{{\rm Re}} & = & \frac{\Re e\,{\cal A}(\gamma^{\ast}+A \to \gamma+A)}
{\Re e \,{\cal A}^{{\rm Born}}(\gamma^{\ast}+A \to \gamma+A)} \,, \nonumber\\
R_{{\rm ampl}}^{{\rm DIS}} & = & \frac{\Im m\,{\cal A}(\gamma^{\ast}+A \to \gamma^{\ast}+A)_{|t=0}}
{\Im m \,{\cal A}^{{\rm Born}}(\gamma^{\ast}+A \to \gamma^{\ast}+A)|_{t=0}} \,,
\label{eq:R_ampl}
\end{eqnarray}
where ${\cal A}(\gamma^{\ast}+A \to \gamma+A)$ is the nuclear DVCS
amplitude of Eq.~(\ref{eq:nucl5});
${\cal A}^{{\rm Born}}(\gamma^{\ast}+A \to \gamma+A)$ is the first
term (Born contribution) of Eq.~(\ref{eq:nucl5});
${\cal A}^{{\rm Born}}(\gamma^{\ast}+A \to \gamma^{\ast}+A)_{|t=0}$ is the first
term  of Eq.~(\ref{eq:nucl6}). Note that  $R_{{\rm ampl}}^{{\rm DIS}}$ is nothing but the 
the ratio of the nuclear to the nucleon inclusive
structure functions,
$R_{{\rm ampl}}^{{\rm DIS}}=F_{2A}(x,Q^2)/[AF_{2N}(x,Q^2)]$.

\begin{figure}[t]
\begin{center}
\epsfig{file=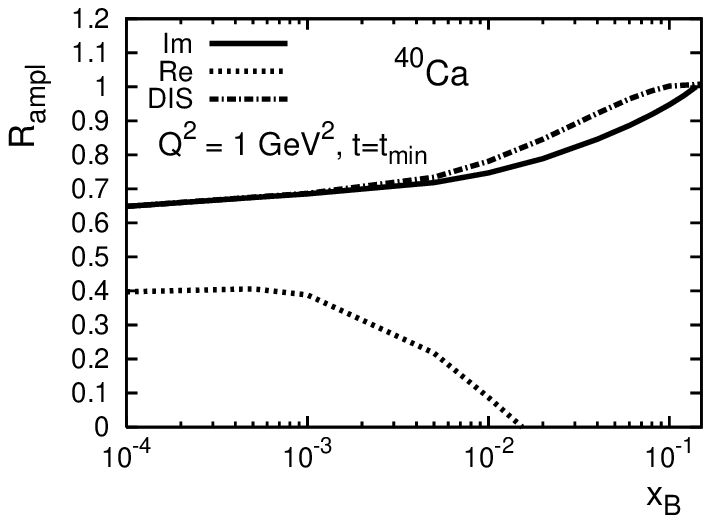,width=8cm,height=8cm}
\epsfig{file=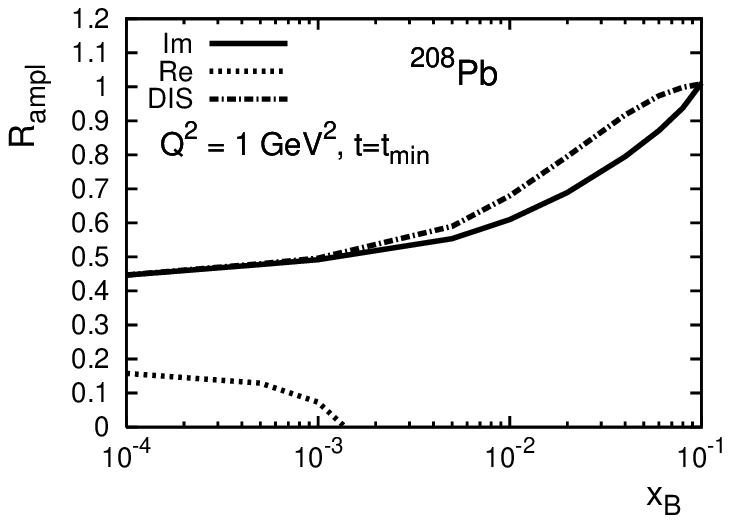,width=8cm,height=8cm}
\caption{The ratios  $R_{{\rm ampl}}^{{\rm Im}}$ (solid),
$R_{{\rm ampl}}^{{\rm Re}}$ (dotted)
 and
$R_{{\rm ampl}}^{{\rm DIS}}$ (dot-dashed) of Eq.~(\ref{eq:R_ampl}) 
at $Q^2=1$ GeV$^2$ and $t=t_{{\rm min}}$
as functions 
of Bjorken $x_B$. 
 The left panel is for $^{40}$Ca; the right panel is for 
$^{208}$Pb.}
\label{fig:nuclear_dvcs_amplitude}
\end{center}
\end{figure}

Figure~\ref{fig:nuclear_dvcs_amplitude} presents the ratios  
$R_{{\rm ampl}}^{{\rm Im}}$ (solid curves),
$R_{{\rm ampl}}^{{\rm Re}}$ (dotted curves) 
 and $R_{{\rm ampl}}^{{\rm DIS}}$ (dot-dashed curves)
 at $Q^2=1$ GeV$^2$ and $t=t_{{\rm min}} \approx -x_B^2 m_N^2$ as functions of $x_B$.
The left panel corresponds to
the nucleus of $^{40}$Ca; the right panel corresponds to $^{208}$Pb.

Let us now discuss the results presented in Fig.~\ref{fig:nuclear_dvcs_amplitude}
in detail. 
The solid and dot-dashed curves coincide for $x_B < 0.01$ and deviate
only slightly for  $0.01 < x_B < 0.1$, which means that the amount of 
nuclear shadowing is the same in the imaginary parts of the DVCS and DIS amplitudes.
This observation agrees with the results obtained within the framework of
a different approach to nuclear GPDs at small-$x_B$, when the latter are modeled using
the align-jet model for the nucleon GPDs and a parameterization of usual nuclear PDFs~\cite{Freund:2003wm}.
Moreover, 
the amount of nuclear shadowing predicted by our 
calculations in the GVMD model
matches very well the leading-twist predictions for $F_{2A}(x,Q^2)/[A F_{2N}(x,Q^2)]$
 made at somewhat higher $Q^2$~\cite{Frankfurt:2003zd}.
This is a consequence of the fact the GVMD model predicts a significant amount
of inclusive diffraction in $\gamma^{\ast} p$ scattering
at all $Q^2$, which controls
 the size of nuclear shadowing in the leading-twist theory of nuclear 
shadowing~\cite{Frankfurt:2003zd}.

As the value of $x_B$ is increased (at fixed $Q^2$),
 the shadowing correction decreases due to 
the decrease of $\sigma_{\rho p}(W)$ and due to the increasingly destructive
role of the $e^{iz_1(k_{\gamma^{\ast}}-k_{V_n})}$ and
$e^{iz_2(k_{V_m}-k_{\gamma})}$ factors in Eq.~(\ref{eq:nucl5}).

For the ratio $R_{{\rm ampl}}^{{\rm Re}}$ of the real parts (dotted curves),
 at small-$x_B$, 
the shadowing correction is 
approximately two times as large as for the ratio of the imaginary parts
because of the fact that $(1-\eta)^2=1-\eta^2-2 i\eta$, see 
Eqs.~(\ref{eq:gvmd_3}) and (\ref{eq:nucl5}).
As $x_B$ increases, the real part of the shadowing correction receives a large contribution
from the $e^{iz_1(k_{\gamma^{\ast}}-k_{V_n})}$ and
$e^{iz_2(k_{V_m}-k_{\gamma})}$ factors, and, as a result, it steadily grows and becomes
larger than the Born contribution.
This behavior of $R_{{\rm ampl}}^{{\rm Re}}$
is similar to that observed in~\cite{Freund:2003wm}. Note, however, that since the
effect of $t_{{\rm min}}$ (the factor $F_A(t_{{\rm min}})$ in the Born term)
was not included in the analysis of~\cite{Freund:2003wm},
the agreement could be coincidental. 

\begin{figure}[t]
\begin{center}
\epsfig{file=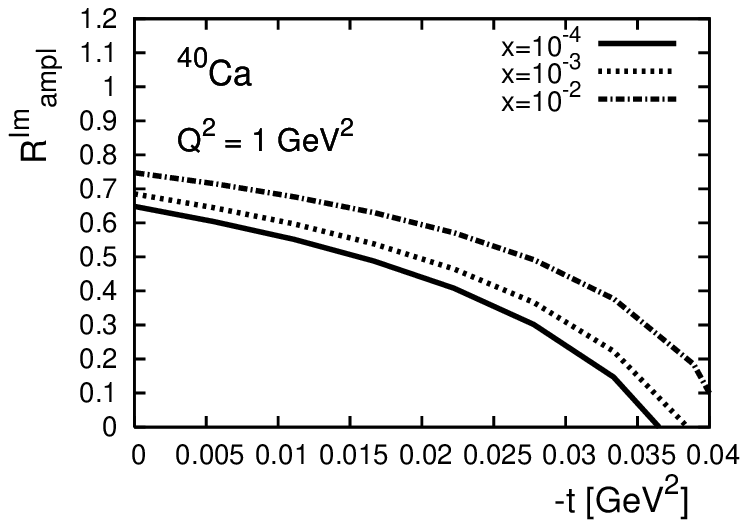,width=8cm,height=8cm}
\epsfig{file=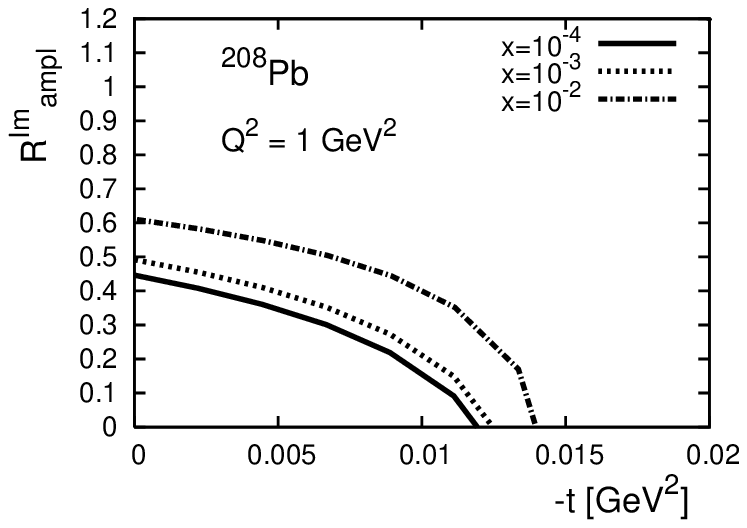,width=8cm,height=8cm}
\caption{The ratio  $R_{{\rm ampl}}^{{\rm Im}}$ at $Q^2=1$ GeV$^2$ 
 as a function of $t$. 
 The left panel is for $^{40}$Ca; the right panel is for 
$^{208}$Pb.}
\label{fig:nuclear_dvcs_amplitude_tdep}
\end{center}
\end{figure}

We also examined nuclear shadowing in DVCS with nuclear targets as a function
of the momentum transfer $t$. Figure~\ref{fig:nuclear_dvcs_amplitude_tdep}
presents $R_{{\rm ampl}}^{{\rm Im}}$ at $Q^2=1$ GeV$^2$ 
as a function of $t$.
Figure~\ref{fig:nuclear_dvcs_amplitude_tdep} demonstrates that the shadowing correction
to the DVCS amplitude has the $t$-dependence which is slower than that of the Born
term.
 As one increases $|t|$, the negative nuclear shadowing correction decreases  
slower than the Born term, which leads to a decrease of $R_{{\rm ampl}}^{{\rm Im}}$.
The ratio $R_{{\rm ampl}}^{{\rm Re}}$ follows the similar trend.

Equation~(\ref{eq:nucl5}) presents the $\gamma^{\ast} A \to \gamma A$ scattering
amplitude as a function of $W$, $Q^2$ and $t$. It also allows for the representation
of the scattering amplitude as a function of $W$, $Q^2$ and $\vec{b}$, where
$\vec{b}$ is the impact parameter conjugate to $\vec{q}_t$,
\begin{eqnarray}
{\cal A}(\gamma^{\ast}+A &\to& \gamma+A)=
\sum_{n,m=0}^{\infty} \frac{e}{f_n}
\frac{M_n^2}{M_n^2+Q^2}  
\frac{e}{f_m} \Bigg[A \int^{\infty}_{-\infty} dz e^{ix_B m_N z} \rho(b,z) \Sigma_{n,m}(W,0) \nonumber\\
&-&\frac{A(A-1)}{2i} \int_{-\infty}^{\infty} dz_1
\int_{z_1}^{\infty} dz_2 \,\rho(b,z_1) 
\rho(b,z_2) e^{iz_1(k_{\gamma^{\ast}}-k_{V_n})}
\nonumber\\
&\times &\tilde{\Sigma}_{n,n^{\prime}}(W,z_1) 
 e^{-\frac{A}{2} \sigma_{\rho p} (W^2)(1-i\eta) \int^{z_2}_{z_1} dz^{\prime}
\rho(b,z^{\prime})}
\tilde{\Sigma}_{n^{\prime},m}(W,z_2) e^{iz_2(k_{V_m}-k_{\gamma})}
\Bigg] \,.
\label{eq:nucl6}
\end{eqnarray} 
In the first term in Eq.~(\ref{eq:nucl6}), we took into account the non-zero
longitudinal momentum transfer, 
$k_{\gamma^{\ast}}-k_{\gamma}=x_B m_N$, see Eq.~(\ref{eq:nucl3}), and
also neglected the $t$-dependence of $\Sigma_{n,m}(t)$ compated to $F_A(t)$.

Using Eq.~(\ref{eq:nucl6}), the nuclear DVCS cross section can be 
expressed in the following compact form
\begin{equation}
\sigma_{{\rm DVCS}}(W,Q^2)
=\frac{1}{4}\int d^2 \vec{b} \, |{\cal A}(\gamma^{\ast}+A \to \gamma+A) |^2 \,.
\label{eq:nucl7}
\end{equation}

In order to quantify predictions of the GVMD model for 
nuclear DVCS cross sections, we introduce the ratio $R_{{\rm cs}}$,
\begin{equation}
R_{{\rm cs}}=
\frac{\sigma_{{\rm DVCS}}(W,Q^2)}{\sigma_{{\rm DVCS}}^{{\rm Born}}(W,Q^2)} \,,
\label{eq:R_cs}
\end{equation}
where the numerator is calculated using Eq.~(\ref{eq:nucl7}) and
the complete expression for the nuclear DVCS amplitude~(\ref{eq:nucl6});
the denominator is calculated using only the first term (Born contribution)
in Eq.~(\ref{eq:nucl6}).

\begin{figure}[t]
\begin{center}
\epsfig{file=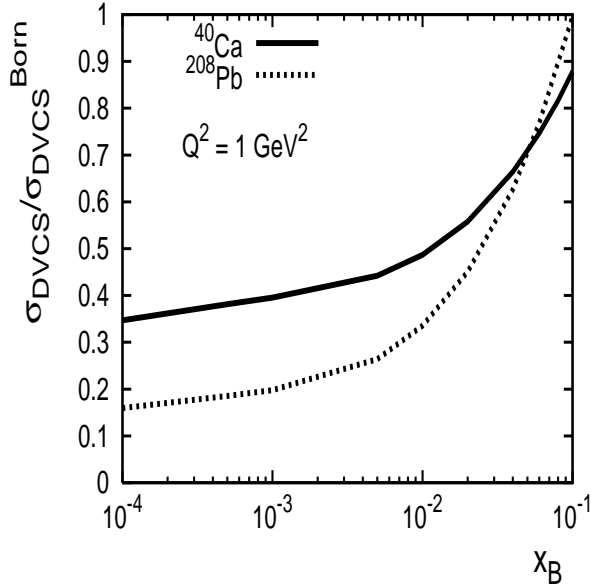,width=8cm,height=8cm}
\caption{The ratio  $R_{{\rm cs}}=\sigma_{{\rm DVCS}} / \sigma_{{\rm DVCS}}^{{\rm Born}}$ of Eq.~(\ref{eq:R_cs}) as a function 
of Bjorken $x_B$ at $Q^2=1$ GeV$^2$. The solid curve 
corresponds to $^{40}$Ca; the dotted curve is for 
$^{208}$Pb.}
\label{fig:nuclear_dvcs_cross_section}
\end{center}
\end{figure}

Predictions of the GVMD model for the ratio $R_{{\rm cs}}$ at $Q^2=1$ GeV$^2$
as a function of $x_B$ are presented in Fig.~\ref{fig:nuclear_dvcs_cross_section}.
The solid curve corresponds to $^{40}$Ca; the dotted curve
corresponds to $^{208}$Pb.

As one can see from Fig.~\ref{fig:nuclear_dvcs_cross_section}, the 
predicted amount of nuclear shadowing at small-$x_B$ is very large.
Since the $t$-dependence of the nuclear DVCS amplitude is very steep,
the dominant contribution to the $t$-integrated cross sections 
entering $R_{{\rm cs}}$ comes from the $t \approx t_{{\rm min}}$ region.
Therefore, the amount of nuclear shadowing for $R_{{\rm cs}}$ is
equal roughly twice the amount of nuclear shadowing for $R_{{\rm ampl}}^{\rm Im}$, 
see Fig.~\ref{fig:nuclear_dvcs_amplitude}.

Finally, we would also like to point out that for nuclear DVCS, 
the ratio of the imaginary parts of the DVCS and DIS amplitudes, see Eq.~(\ref{eq:R_1}) and Fig.~\ref{fig:Schoeffel}, is quite similar to the free nucleon case.
This is a mere consequence of the fact that the structure of the
$Q^2$-dependence of the ratio is essentially the same in the DVCS on
the nucleon and on nuclei.

\section{Conclusions and Discussion}
\label{sec:conclusions}

We considered Deeply Virtual Compton Scattering (DVCS) on
nucleons and nuclei in the framework of generalized vector
meson dominance (GVMD) model. We extended the original GVMD model,
which was applied to forward amplitudes of 
Deep Inelastic Scattering (DIS), to the non-forward $t \neq 0$ case.
We introduced the $W$-dependence of the DVCS amplitude through the 
$W$-dependence of the elementary vector meson-nucleon amplitude, which 
was taken to be proportional to $W^{0.2}$ at high-$W$.

We compared our predictions to the HERA data on DVCS on the proton
 with the following results. The GVMD model describes well
the dependence of the DVCS cross section on $Q^2$, $W$ (at  $Q^2$=4 GeV$^2$) and $t$.
At $Q^2$=8 GeV$^2$, the $W$-dependence of the 
cross section is somewhat underestimated, which 
can be interpreted as due to the onset of the 
hard regime beyond the soft dynamics of the GVMD model.

We estimated the relative contribution of $1/Q^2$-corrections
to the DVCS amplitude and the DVCS cross section. 
We found that these corrections are large:
the contribution of the $1/Q^2$-corrections to the DVCS amplitude at 
$t=t_{{\min}}$ is 20\% at $Q^2=2$ GeV$^2$, 11\% at $Q^2=4$ GeV$^2$
and 6\% at $Q^2=8$ GeV$^2$;
the contribution of the $1/Q^2$-corrections 
 to the $t$-integrated DVCS cross
section is 56\% at $Q^2=2$ GeV$^2$, 32\% at $Q^2=4$ GeV$^2$
and 17\% at $Q^2=8$ GeV$^2$.

We also made predictions for the DVCS amplitude and the DVCS cross section on nuclear targets, which are relevant for the physics program of the future Electron-Ion Collider.
We predicted significant nuclear shadowing, which matches well predictions 
of the leading-twist nuclear shadowing in DIS on nuclei.

Our analysis allows us to argue that the GVMD model provides a reliable
parameterization of the DVCS amplitude and the DVCS cross section
with nucleons and nuclei in a wide range of
kinematics.
At fixed values of $Q^2$, which should not be too large, $Q^2 \lesssim 5$ GeV$^2$,
the GVMD model is applicable starting from $W=2$ GeV (JLab), towards
$W \approx 80$ GeV (HERA) and beyond (real photons at the LHC).
 At fixed $W$, the GVMD model is applicable from the photoproduction limit
up to the values of $Q^2$, where perturbative QCD can already be used, 
$0 \leq Q^2 \lesssim 5$ GeV$^2$.
In addition, due to the correct $1/Q^2$-scaling, predictions of the GVMD model
can be extrapolated to much higher values of $Q^2$ such that the range of
applicability of the GVMD model becomes very wide,
$0 \leq Q^2 <80$ GeV$^2$.
The model can be applied for a wide range of $t$:
$0 < ŧ|t| < 1$ GeV$^2$.

\begin{acknowledgements}

The authors would like to thank C.~Weiss for suggesting the research 
topic of  the present paper and for useful discussions.
We also thank M.~Strikman for reading the manuscript and useful comments.

The work has been partially supported by the Verbundforschung ("Hadronen und Kerne") of the BMBF and by the Transregio/SFB Bonn-Bochum-Giessen.

{\bf Notice}: Authored by Jefferson Science Associates, LLC under U.S. DOE Contract No. DE-AC05-06OR23177. The U.S. Government retains a non-exclusive, paid-up, irrevocable, world-wide license to publish or reproduce this manuscript for U.S. Government purposes.

\end{acknowledgements}


\begin{thebibliography}{99}

\bibitem{Mueller:1998fv}
  D.~Mueller, D.~Robaschik, B.~Geyer, F.~M.~Dittes and J.~Horejsi,
  Fortsch.\ Phys.\  {\bf 42}, 101 (1994)
  [arXiv:hep-ph/9812448].

\bibitem{Ji:1996nm}
  X.~D.~Ji,
  Phys.\ Rev.\  D {\bf 55}, 7114 (1997).

\bibitem{Ji:1998pc}
  X.~D.~Ji,
  J.\ Phys.\ G {\bf 24}, 1181 (1998)
  [arXiv:hep-ph/9807358].

\bibitem{Radyushkin:1996nd}
  A.~V.~Radyushkin,
  Phys.\ Lett.\  B {\bf 380}, 417 (1996)
  [arXiv:hep-ph/9604317].


\bibitem{Radyushkin:1997ki}
  A.~V.~Radyushkin,
  Phys.\ Rev.\  D {\bf 56}, 5524 (1997).

\bibitem{Radyushkin:2000uy}
  A.~V.~Radyushkin,
  arXiv:hep-ph/0101225.

\bibitem{Collins:1998be}
  J.~C.~Collins and A.~Freund,
  Phys.\ Rev.\  D {\bf 59}, 074009 (1999).

\bibitem{Collins:1996fb}
  J.~C.~Collins, L.~Frankfurt and M.~Strikman,
  Phys.\ Rev.\  D {\bf 56}, 2982 (1997).

\bibitem{Brodsky:1994kf}
  S.~J.~Brodsky, L.~Frankfurt, J.~F.~Gunion, A.~H.~Mueller and M.~Strikman,
  Phys.\ Rev.\  D {\bf 50}, 3134 (1994).

\bibitem{Goeke:2001tz}
  K.~Goeke, M.~V.~Polyakov and M.~Vanderhaeghen,
  Prog.\ Part.\ Nucl.\ Phys.\  {\bf 47}, 401 (2001)
  [arXiv:hep-ph/0106012].

\bibitem{Diehl:2000xz}
  M.~Diehl, T.~Feldmann, R.~Jakob and P.~Kroll,
  Nucl.\ Phys.\  B {\bf 596}, 33 (2001)
  [Erratum-ibid.\  B {\bf 605}, 647 (2001)]
  [arXiv:hep-ph/0009255].


\bibitem{Belitsky:2001ns}
  A.~V.~Belitsky, D.~Mueller and A.~Kirchner,
  Nucl.\ Phys.\  B {\bf 629}, 323 (2002)
  [arXiv:hep-ph/0112108].

\bibitem{Diehl:2003ny}
  M.~Diehl,
  Phys.\ Rept.\  {\bf 388}, 41 (2003)
  [arXiv:hep-ph/0307382].

\bibitem{Belitsky:2005qn}
  A.~V.~Belitsky and A.~V.~Radyushkin,
  Phys.\ Rept.\  {\bf 418}, 1 (2005)
  [arXiv:hep-ph/0504030].

\bibitem{Berger:2001xd}
  E.~R.~Berger, M.~Diehl and B.~Pire,
  Eur.\ Phys.\ J.\  C {\bf 23}, 675 (2002)
  [arXiv:hep-ph/0110062].

\bibitem{Berger:2001zn}
  E.~R.~Berger, M.~Diehl and B.~Pire,
  Phys.\ Lett.\  B {\bf 523}, 265 (2001)
  [arXiv:hep-ph/0110080].

\bibitem{Diehl:2000uv}
  M.~Diehl, T.~Gousset and B.~Pire,
  Phys.\ Rev.\  D {\bf 62}, 073014 (2000)
  [arXiv:hep-ph/0003233].



\bibitem{Mankiewicz:1998kg}
  L.~Mankiewicz, G.~Piller and A.~Radyushkin,
  Eur.\ Phys.\ J.\  C {\bf 10}, 307 (1999)
  [arXiv:hep-ph/9812467].

\bibitem{Frankfurt:1999fp}
  L.~L.~Frankfurt, P.~V.~Pobylitsa, M.~V.~Polyakov and M.~Strikman,
  Phys.\ Rev.\  D {\bf 60}, 014010 (1999).

\bibitem{Airapetian:2001yk}
  A.~Airapetian {\it et al.}  [HERMES Collaboration],
  Phys.\ Rev.\ Lett.\  {\bf 87}, 182001 (2001).

\bibitem{Airapetian:2006zr}
  A.~Airapetian {\it et al.}  [HERMES Collaboration],
  Phys.\ Rev.\  D {\bf 75}, 011103 (2007).

\bibitem{Ye:2006gza}
  Z.~Ye  [HERMES Collaboration],
  arXiv:hep-ex/0606061.

\bibitem{Stepanyan:2001sm}
  S.~Stepanyan {\it et al.}  [CLAS Collaboration],
  Phys.\ Rev.\ Lett.\  {\bf 87}, 182002 (2001).

\bibitem{Munoz Camacho:2006hx}
  C.~Munoz Camacho {\it et al.}  [Jefferson Lab Hall A Collaboration],
  Phys.\ Rev.\ Lett.\  {\bf 97}, 262002 (2006).

\bibitem{Girod:2007jq}
  F.~X.~Girod {\it et al.}  [CLAS Collaboration],
  arXiv:0711.4805 [hep-ph].

\bibitem{Fraas:1974gh}
  H.~Fraas, B.~J.~Read and D.~Schildknecht,
  Nucl.\ Phys.\  B {\bf 86}, 346 (1975).

\bibitem{Ditsas:1975vd}
  P.~Ditsas, B.~J.~Read and G.~Shaw,
  Nucl.\ Phys.\  B {\bf 99}, 85 (1975).

\bibitem{Shaw:1993gx}
  G.~Shaw,
  Phys.\ Rev.\  D {\bf 47}, 3676 (1993).


\bibitem{Frankfurt:1997zk}
  L.~Frankfurt, V.~Guzey and M.~Strikman,
  Phys.\ Rev.\  D {\bf 58}, 094039 (1998).

\bibitem{Chekanov:2003ya}
  S.~Chekanov {\it et al.}  [ZEUS Collaboration],
  Phys.\ Lett.\  B {\bf 573}, 46 (2003)
  [arXiv:hep-ex/0305028].

\bibitem{Aktas:2005ty}
  A.~Aktas {\it et al.}  [H1 Collaboration],
  Eur.\ Phys.\ J.\  C {\bf 44}, 1 (2005)
  [arXiv:hep-ex/0505061].

\bibitem{Kivel:2000fg}
  N.~Kivel, M.~V.~Polyakov and M.~Vanderhaeghen,
  Phys.\ Rev.\  D {\bf 63}, 114014 (2001).

\bibitem{Freund:2003qs}
  A.~Freund,
  Phys.\ Rev.\  D {\bf 68}, 096006 (2003).

\bibitem{Radyushkin:2000ap}
  A.~V.~Radyushkin and C.~Weiss,
  Phys.\ Rev.\  D {\bf 63}, 114012 (2001).

\bibitem{Frankfurt:2003zd}
  L.~Frankfurt, V.~Guzey and M.~Strikman,
  Phys.\ Rev.\  D {\bf 71}, 054001 (2005).

\bibitem{Feynman}
R.~P.~Feynman, {\it Photon-Hadron Interactions} 
(Benjamin, Reading, 1972).

\bibitem{Bauer:1977iq}
  T.~H.~Bauer, R.~D.~Spital, D.~R.~Yennie and F.~M.~Pipkin,
  Rev.\ Mod.\ Phys.\  {\bf 50}, 261 (1978)
  [Erratum-ibid.\  {\bf 51}, 407 (1979)].

\bibitem{Fujikawa:1972ux}
  K.~Fujikawa,
  Phys.\ Rev.\  D {\bf 4}, 2794 (1971).


\bibitem{Gribov:1968gs}
  V.~N.~Gribov,
  Sov.\ Phys.\ JETP {\bf 30} (1970) 709
  [Zh.\ Eksp.\ Teor.\ Fiz.\  {\bf 57} (1969) 1306].


\bibitem{Fraas:1971hk}
  H.~Fraas and D.~Schildknecht,
  Phys.\ Lett.\  B {\bf 35}, 72 (1971).


\bibitem{Dar:1971wh}
  A.~Dar,
  Annals Phys.\  {\bf 65}, 324 (1971).

\bibitem{Yao:2006px}
  W.~M.~Yao {\it et al.}  [Particle Data Group],
  J.\ Phys.\ G {\bf 33}, 1 (2006).


\bibitem{Aid:1996bs}
  S.~Aid {\it et al.}  [H1 Collaboration],
  Nucl.\ Phys.\  B {\bf 463}, 3 (1996)
  [arXiv:hep-ex/9601004].

\bibitem{Breitweg:1997ed}
  J.~Breitweg {\it et al.}  [ZEUS Collaboration],
  Eur.\ Phys.\ J.\  C {\bf 2}, 247 (1998)
  [arXiv:hep-ex/9712020].

\bibitem{Adloff:1999kg}
  C.~Adloff {\it et al.}  [H1 Collaboration],
  Eur.\ Phys.\ J.\  C {\bf 13}, 371 (2000)
  [arXiv:hep-ex/9902019].


\bibitem{Pautz:1997eh}
  A.~Pautz and G.~Shaw,
  Phys.\ Rev.\  C {\bf 57}, 2648 (1998)
  [arXiv:hep-ph/9710235].

\bibitem{Chekanov:2001gw}
  S.~Chekanov {\it et al.}  [ZEUS Collaboration],
  Nucl.\ Phys.\  B {\bf 627}, 3 (2002)
  [arXiv:hep-ex/0202034].

\bibitem{Gribov:1968uy}
  V.~N.~Gribov and A.~A.~Migdal,
  Sov.\ J.\ Nucl.\ Phys.\  {\bf 8}, 583 (1969)
  [Yad.\ Fiz.\  {\bf 8}, 1002 (1968)].


\bibitem{Schoeffel:2007dt}
  L.~Schoeffel,
  arXiv:0706.3488 [hep-ph].

\bibitem{Freund:2002qf}
  A.~Freund, M.~McDermott and M.~Strikman,
  Phys.\ Rev.\  D {\bf 67}, 036001 (2003).


\bibitem{Hand:1963bb}
  L.~N.~Hand,
  Phys.\ Rev.\  {\bf 129}, 1834 (1963).




\bibitem{Piller:1999wx}
  G.~Piller and W.~Weise,
  Phys.\ Rept.\  {\bf 330}, 1 (2000)
  [arXiv:hep-ph/9908230].

\bibitem{DeJager:1987qc}
  C.~W.~De Jager, H.~De Vries and C.~De Vries,
  Atom.\ Data Nucl.\ Data Tabl.\  {\bf 36}, 495 (1987).

\bibitem{Freund:2003wm}
  A.~Freund and M.~Strikman,
  Phys.\ Rev.\  C {\bf 69}, 015203 (2004).

\end{thebibliography}
\end{document}